\documentclass[pra,aps,twocolumn]{revtex4}
\usepackage{amsmath,graphicx,epsfig}

\def\prn#1{{\left(#1\right)}}

\def\sbrk#1{{\left[#1\right]}}
\def\abrk#1{{\langle#1\rangle}}

\def\ket#1{{|#1\rangle}}
\def\bra#1{{\langle#1|}}

\def\fig_width{3. in} % width of single column figure in PR
\draft

\newlength{\defbaselineskip}
\newcommand{\setlinespacing}[1]%
           {\setlength{\baselineskip}{#1 \defbaselineskip}}

\begin{document}

\title{AC Stark shift noise in QND measurement arising from quantum fluctuations of light polarization} %==================================

\author{M. Auzinsh}
\affiliation{Department of Physics, University of Latvia, 19 Rainis blvd, Riga, LV-1586,
Latvia}
\author{D. Budker}
\email{budker@socrates.berkeley.edu} \affiliation{Department of Physics, University of
California, Berkeley, CA 94720-7300} \affiliation{Nuclear Science Division, Lawrence
Berkeley National Laboratory, Berkeley CA 94720}
\author{D. F. Kimball}
\affiliation{Department of Physics, University of California, Berkeley, CA 94720-7300}
\author{S. M. Rochester}
\affiliation{Department of Physics, University of California, Berkeley, CA 94720-7300}
\author{J. E. Stalnaker}
\affiliation{Department of Physics, University of California, Berkeley, CA 94720-7300}
\author{A. O. Sushkov}
\affiliation{Department of Physics, University of California, Berkeley, CA 94720-7300}
\author{V. V. Yashchuk}
\affiliation{Advanced Light Source Division, Lawrence Berkeley
National Laboratory, Berkeley CA 94720}

\begin{abstract}
In a recent letter [Auzinsh {\it{et. al.}} (physics/0403097)] we have analyzed the noise
properties of an idealized atomic magnetometer that utilizes spin squeezing induced by a
continuous quantum nondemolition measurement. Such a magnetometer measures spin
precession of $N$ atomic spins by detecting optical rotation of far-detuned probe light.
Here we consider maximally squeezed probe light, and carry out a detailed derivation of
the contribution to the noise in a magnetometric measurement due to the differential AC
Stark shift between Zeeman sublevels arising from quantum fluctuations of the probe
polarization.
\end{abstract} 

\date{\today}

\maketitle

In this companion note to Ref.~\cite{OurQND}, we explicitly perform a calculation of the
effect of quantum fluctuations of the probe light polarization on the atomic spins. (All
references to equations not preceded by ``A'' refer to equations from
Ref.~\cite{OurQND}.)

For nominally $y$-polarized light there is a small, random admixture of circular
polarization caused by vacuum fluctuations in the orthogonal polarization. Such random
circular polarization causes a differential AC Stark shift between the Zeeman sublevels.
Consequently, atoms initially polarized along $x$ will precess in the $xy$-plane by an
angle $\alpha_{xy}$.

As discussed in Ref.~\cite{OurQND} [Eq.~(19)], the quantum fluctuations of the probe
light polarization can be described using the ellipticity operator introduced in
Ref.~\cite{OurSqueezing}:
\begin{align}
\hat{\epsilon} = \frac{{\mathcal{E}_0}}{2iE}\prn{\hat{a}_x-\hat{a}_x^\dag}~.
\end{align}
The quantum fluctuations of the ellipticity can be calculated from Eq.~(20)
\begin{align}
\delta\epsilon = \sqrt{\abrk{\hat{\epsilon}^2}-\abrk{\hat{\epsilon}}^2}~. \nonumber
\end{align}
Assuming that the $x$-polarized field is the unsqueezed vacuum $\ket{0}$, for which
\begin{align}
\hat{a}_x \ket{0} & = 0\ket{0}~, \label{Eq:AnnihilationOnVacuum}\\
\bra{0}\hat{a}_x^\dag & = 0\bra{0}~, \label{Eq:CreationOnVacuum}
\end{align}
we have
\begin{align}
\abrk{\hat{\epsilon}} = \frac{{\mathcal{E}_0}}{2iE}
\bra{0}\prn{\hat{a}_x-\hat{a}_x^\dag}\ket{0} = 0~,
\end{align}
and also
\begin{align}
\abrk{\hat{\epsilon}^2} & =
-\frac{{\mathcal{E}_0}^2}{4E^2}\bra{0}\prn{\hat{a}_x-\hat{a}_x^\dag}^2\ket{0} \\
& = -\frac{{\mathcal{E}_0}^2}{4E^2}
\bra{0}\prn{\hat{a}_x\hat{a}_x-\hat{a}_x\hat{a}_x^\dag-\hat{a}_x^\dag\hat{a}_x+\hat{a}_x^\dag\hat{a}_x^\dag}\ket{0}
\label{Eq:Step1InEllipticityNoiseDerivation}\\
& = \frac{{\mathcal{E}_0}^2}{4E^2}\bra{0}\prn{\hat{a}_x\hat{a}_x^\dag}\ket{0}~,
\label{Eq:Step2InEllipticityNoiseDerivation}
\end{align}
where in Eq.~\eqref{Eq:Step1InEllipticityNoiseDerivation} all terms except the second one
are zero by the properties \eqref{Eq:AnnihilationOnVacuum} and
\eqref{Eq:CreationOnVacuum}.  By employing the relation $\hat{a}_x\hat{a}_x^\dag = 1 +
\hat{a}_x^\dag\hat{a}_x$ (derived from the commutation relation
$\sbrk{\hat{a}_x,\hat{a}_x^\dag}=1$) to Eq.~\eqref{Eq:Step2InEllipticityNoiseDerivation},
we obtain
\begin{align}
\abrk{\hat{\epsilon}^2} = \frac{{\mathcal{E}_0}^2}{4E^2} \sim \frac{1}{N_{ph}}~.
\end{align}
Thus we find
\begin{align}
\boxed{\delta\epsilon \sim \frac{1}{\sqrt{N_{ph}}}~,} \label{Eq:EllipNoise}
\end{align}
as stated in Eq.~(20).

Equation (21) from Ref.~\cite{OurQND} gives the resulting differential AC Stark shift of
the ground state magnetic sublevels:
%--------------------------------------------------------------------------
\begin{equation}
\delta\Delta_{ac} = \frac{d^2E^2}{\Delta}\delta\epsilon~. \nonumber
\end{equation}
%--------------------------------------------------------------------------
>From Eqs.~(6), (20), and (21) we obtain the expression
\begin{align}
\delta\Delta_{ac} = \frac{d^2}{\Delta}\frac{N_{ph}}{\lambda A
\tau}\frac{1}{\sqrt{N_{ph}}}~, \label{Eq:ACstark}
\end{align}
The differential Stark shift causes the atomic polarization vector to precess by a random
angle in the $xy$-plane, and after time $\tau$ it has rotated by
\begin{align}
\alpha_{xy} = \tau\delta\Delta_{ac}~. \label{Eq:PrecBasic}
\end{align}
Thus we have, by substituting Eq.~\eqref{Eq:ACstark} into Eq.~\eqref{Eq:PrecBasic} and
employing the relation $d^2=\lambda^3\Gamma_0$,
\begin{align}
\alpha_{xy} = \frac{\lambda^2}{A}\frac{\Gamma_0}{\Delta}\sqrt{N_{ph}}~.
\label{Eq:AlphaXYunopt}
\end{align}

Equation (9) gives the optimum number of probe photons to minimize the noise in the
magnetic field measurement:
\begin{align}
N_{ph}^{\rm(opt)} = \frac{1}{\sqrt{N}} \prn{\frac{\Delta}{\Gamma_0}}^2
\prn{\frac{A}{\lambda^2}}^{3/2}~,
\end{align}
which, when substituted for $N_{ph}$ in Eq.~\eqref{Eq:AlphaXYunopt}, yields Eq.~(22):
\begin{align}
\boxed{\alpha_{xy} = \frac{1}{N^{1/4}}\left(\frac{A}{\lambda^2}\right)^{-1/4}.} \nonumber
\end{align}

What if squeezed probe light is used?  In order to gain an advantage in polarimetric
sensitivity, one must employ a local oscillator and heterodyne detection.  Then the noise
level is different depending on the phase of the local oscillator, and can in fact be
less than shot noise. For example, if the squeezing technique discussed in
Ref.~\cite{OurSqueezing} is employed, the annihilation operator for squeezed light
($\hat{a}_x^s$) is given by
\begin{align}
\hat{a}_x^s = \prn{e^{i\chi}-ig\ell \cos\chi}\hat{a}_x~, \label{Eq:SqueezedOperator}
\end{align}
where $g\ell$ is a measure of the degree of squeezing, $\chi$ is the phase of the local
oscillator, and the operator $\hat{a}_x$ acts on the unsqueezed light states, in our case
the vacuum $\ket{0}$.  The quantum fluctuations of the squeezed probe light polarization
are described by
\begin{align}
\abrk{\delta\epsilon}^2 = \frac{1}{N_{ph}}\prn{1-g\ell \sin \chi \cos \chi + g^2\ell^2
\cos^2\chi}~, \label{Eq:SqueezedLightEllipNoise}
\end{align}
where Eq.~\eqref{Eq:SqueezedLightEllipNoise} can be derived in the same manner that
Eq.~\eqref{Eq:EllipNoise} was obtained by employing the expression
\eqref{Eq:SqueezedOperator} for the annihilation operator (as is done explicitly in
Ref.~\cite{OurSqueezing}).

Assuming there is no absorption in the squeezing medium (which would degrade the amount
of squeezing), for maximally squeezed light with the phase of the local oscillator $\chi$
chosen to produce minimum noise (see Ref.~\cite{OurSqueezing}), we find that the noise in
a polarimetric measurement of the optical rotation angle $\varphi$ is
\begin{align}
\delta\varphi \sim \frac{1}{g^2\ell^2}~.
\end{align}
In order to satisfy the Heisenberg limit this requires that
\begin{align}
g^2\ell^2 = N_{ph}~.
\end{align}

However, atoms interact with all possible phases of the squeezed light, and thus the
relevant ellipticity fluctuations for calculating the AC Stark shift effects for squeezed
light are obtained by averaging over all phases:
\begin{align}
\abrk{\delta\epsilon}^2 & = \frac{1}{2\pi} \int_0^{2\pi}\prn{\frac{1}{N_{ph}}-\frac{\sin
\chi \cos \chi}{\sqrt{N_{ph}}} + \cos^2\chi}d\chi \nonumber \\
& = \frac{1}{N_{ph}} + \frac{1}{2} \approx 1~.
\end{align}
This justifies the statement in Ref.~\cite{OurQND} that $\delta\epsilon \sim 1$ for
maximally squeezed light.

\end{document}